# Robust flat bands in twisted trilayer graphene quasicrystals

**Author list:** Chen-Yue Hao[1,2,‡], Zhen Zhan[3,‡,†], Pierre A. Pantaleón[3], Jia-Qi He[1,2], Ya-Xin Zhao[1,2], Kenji Watanabe[4], Takashi Taniguchi[5], Francisco Guinea[3,6,7], Lin He[1,2,†]

**Affiliations:**

[1]Center for Advanced Quantum Studies, Department of Physics, Beijing Normal University, Beijing, 100875, China

[2]Key Laboratory of Multiscale Spin Physics, Ministry of Education, Beijing, 100875, China

[3]IMDEA Nanoscience, Faraday 9, 28049 Madrid, Spain

[4]Research Center for Functional Materials, National Institute for Materials Science, Tsukuba, Japan

[5]International Center for Materials Nanoarchitectonics, National Institute for Materials Science, Tsukuba, Japan

[6]Donostia International Physics Center, Paseo Manuel de Lardizábal 4, 20018 San Sebastián, Spain

[7]Ikerbasque, Basque Foundation for Science, 48009 Bilbao, Spain

‡These authors contributed equally

†Correspondence and requests for materials should be addressed to Zhen Zhan (e-mail: zhenzhanh@gmail.com) and Lin He (e-mail: helin@bnu.edu.cn).

**Abstract:**

**Moiré structures formed by twisting three layers of graphene with two independent twist angles present an ideal platform for studying correlated**

**quantum phenomena, as an infinite set of angle pairs is predicted to exhibit flat bands. Moreover, the two mutually incommensurate moiré patterns among the twisted trilayer graphene (TTG) can form highly tunable moiré quasicrystals. This enables us to extend correlated physics in periodic moiré crystals to quasiperiodic systems. However, direct local characterization of the structure of the moiré quasicrystals and of the resulting flat bands are still lacking, which is crucial to fundamental understanding and control of the correlated moiré physics. Here, we demonstrate the existence of flat bands in a series of TTGs with various twist angle pairs and show that the TTGs with different magic angle pairs are strikingly dissimilar in their atomic and electronic structures. The lattice relaxation and the interference between moiré patterns are highly dependent on the twist angles. Our direct spatial mappings, supported by theoretical calculations, reveal that the localization of the flat bands exhibits distinct symmetries in different regions of the moiré quasicrystals.**

Two-dimensional (2D) moiré superlattices are ideal platforms to explore strongly correlated phases due to the presence of nearly flat bands, where the Coulomb interactions dominate over the kinetic energy[1-3]. This can be achieved by controlling the twist angle and was first observed in twisted bilayer graphene (TBG) with a special twist angle ~ 1.1° (the so-called magic angle)[4-9]. The observed correlated insulator states, superconductivity, quantum anomalous Hall effect and ferromagnetic Chern insulators in the magic-angle TBG have sparked substantial research on various 2D moiré systems[10-23], such as twisted trilayer graphene (TTG)[10-14], twisted double-bilayer graphene[15,16], and twisted transition metal dichalcogenides[19-23], aiming at a deeper understanding of the underlying moiré physics.

Recently, particular attentions have been focused on the TTG[10-14,24-31], which is predicted to exhibit robust flat bands in an infinite set of magic-angle pairs[31-35]. Such a flexible system provides a more versatile platform for exploring the flat-bands-related novel correlated phases. The TTG is characterized by two independent twist angles, $\theta_{TM}$ and $\theta_{MB}$, which correspond to the rotation angles of top (T) to middle (M) layers, and M to bottom (B) layers, respectively. For the general case where the ratio $\theta_{TM}/\theta_{MB}$ is away from one, we have two incommensurate moiré periodicities, which allows us to realize a rich class of distinct moiré quasicrystals. Very recently[25], superconductivity was observed in a TTG moiré quasicrystal with an angle pair ($\theta_{TM} = 1.41°, \theta_{MB} = -1.88°$), suggesting that the strict periodic structure is not a prerequisite for correlated physics. Therefore, the TTG is a new and convenient family for the study of novel correlated phenomena in moiré quasiperiodic systems. However, up to now, microscopic knowledge of the structure and of the electronically flat bands of the moiré quasicrystals is still lacking. For example, little is known about the structure reconstruction of the moiré quasicrystals and a direct measurement of the flat bands of distinct moiré quasicrystals have not been performed yet. In this work, we present local characterization of a series of the TTG moiré quasicrystals via scanning tunnelling microscopy (STM) and scanning tunnelling spectroscopy (STS). We observe spectroscopic signatures of the flat bands in a collection of TTGs, which provide direct confirmation of the theoretically predicted magic phase in such moiré quasicrystals. The spatial distributions of the flat bands in different regions of the TTGs are directly imaged in the STM measurements, showing distinct crystal symmetries. Our experiments supported by theoretical calculations show that the moiré pattern reconstruction and interference in the TTGs strongly depend on the twist angles.

A schematic of our experimental set-up in the STM measurement is shown in Fig. 1a. The TTG samples with controlled twist angles ($\theta_{TM}$, $\theta_{MB}$) were fabricated by a dry-transfer technique on a hexagonal-boron nitride (hBN) substrate (Methods). The graphene layers are intentionally misaligned with the hBN to alleviate substrate effects. For our discussion, we assume that the two angles in the TTG are approximately commensurate, $\theta_{TM} \approx m\theta_0$, $\theta_{MB} \approx n\theta_0$, where m and n are coprime integers. Here $\theta_0$ is approximated to give a period $l_{\theta_0} \approx a/[2\sin(\theta_0/2)]$ if we neglect the angle between $\theta_{TM}$ and $\theta_{MB}$, being $a$ the lattice constant of graphene, as reported in refs.[31,32]. It has been demonstrated theoretically that the flat bands appear in many different angle ratio ($\theta_{TM}/\theta_{MB}$) with an infinite set of $\theta_0$, i.e. a magic phase[32,33]. For example, the continuum calculations of angle pairs ($\theta_{TM}$, $\theta_{MB}$) with {m=1, n=-2} and {m=2, n=-3} predict an infinite collection of magic $\theta_0$ that possess flat bands (Fig. 1b). A direct experimental confirmation of the predicted magic angle phase is highly demanded. Furthermore, theoretical investigations have shown that the lattice relaxation plays a crucial role in the TTG[29,30]. In the case of $\theta_{TM}/\theta_{MB} \approx \pm 1$, the $\theta_0$ is tiny with $\theta_0 \ll \theta_{TM}, \theta_{MB}$. Subsequently, structure reconstruction leads to the formation of a single-moiré structure over a large area with a length $l_{\theta_0}$[31]. In this work, our focus is on the quasicrystal cases where the ratio $\theta_{TM}/\theta_{MB}$ is away from one, for instance, $\frac{\theta_{TM}}{\theta_{MB}} \approx -\frac{1}{2}, -\frac{1}{3}, -\frac{2}{3}, -\frac{3}{4}....$ In this context, both $\theta_{TM}$ and $\theta_{MB}$ not only give rise to moiré periods $\lambda_{TM}$ and $\lambda_{MB}$ respectively, and quasiperiodicity $l_{\theta_0}$, but also results in a super-long-range moiré-of-moiré (MoM) periodicity[32] with length $l_{MoM} = a/[2\sin(\theta_0/2)]^2$. The MoM consists of three distinct types of stacking configurations: AA-MoM, BA/AB-MoM, DW-MoM, as schematically shown in Fig. 1c for the case $\frac{\theta_{TM}}{\theta_{MB}} \approx -\frac{1}{2}$. The incommensurability of the two periodicities $\lambda_{TM}$ and $\lambda_{MB}$ produces moiré-scale quasiperiodicity[25] with length of $l_{\theta_0}$, as schematically shown in Fig. 1d. A distinctive feature of the MoM is that the $C_6$ symmetry of stacking orders is preserved only in the AA-MoM region, while it is broken in other stacking regions (Fig. 1d). However, a direct local characterization of the MoM and lattice reconstruction in TTG quasicrystals has so far been lacking.

In this work, five TTG samples with different twist angle pairs ($\theta_{TM}$, $\theta_{MB}$) are studied. The detailed structural information of the five TTG samples are listed in table 1. Figure 2a shows a representative STM image of a TTG with the twist angle pair (2.85°, -1.9°), i.e., the sample 1 with $\theta_0 = 0.95°$, m = 3, and n = -2. The moiré superlattices with $\lambda_{TM}$ = 4.95 nm, $\lambda_{MB}$ = 7.42 nm, and the supermodulation $l_{\theta_0}$ = 14.84 nm can be clearly identified in both the STM image and its fast Fourier transform (FFT) image (Fig. 2b). The unit cell of the MoM is estimated as $l_{MoM} = a/[2\sin(\theta_0/2)]^2$ = 894.9 nm, which is too large to be directly imaged in the STM measurements. However, we can identify the AA-MoM region of the MoM according to the local $C_6$ symmetry of the moiré structures, as summarized in Figs. 2c-2f. Figures 2c and 2e show two representative STM images recorded in the central area of the AA-MoM region and about 100 nm away from the AA-MoM region, respectively. Although the moiré superlattices $\lambda_{TM}$, $\lambda_{MB}$, and $l_{\theta_0}$ are identical in the two STM images, they exhibit quite different local symmetry of the moiré patterns: the $C_6$ symmetry is preserved in Fig. 2c, whereas it is broken in Fig. 2e. Such a result is verified from the topography profiles along $L_1$, $L_2$ and $L_3$, as shown in Figs. 2d and 2f. In the central area of the AA-MoM region, the three profiles exhibit the same period $\lambda_{TM}$ = 4.95 nm with a supermodulation $l_{\theta_0}$ = 14.84 nm (Fig. 2d). However, in the area away from the AA-MoM region, the profile lines $L_2$ and $L_3$ exhibit noticeably different features in both periodicity and magnitude from that along the $L_1$, as shown in Fig. 2f.

In our experiment, the same STM measurements are carried out in all the studied TTG quasicrystals and similar features are observed in the local moiré symmetry of the MoM. Figure 3 shows representative results obtained in two TTG samples, i.e., the sample 2 with a twist angle pair (0.54°, -1.08°) and $\theta_0$ = 0.54°, and the sample 3 with a twist angle pair (0.88°, -1.76°) and $\theta_0$ = 0.88°. First, we can identify the studied regions of the MoM according to both the local moiré symmetry in the STM images, as shown in Figs. 3a and 3e, and the high-symmetry profile lines in Figs. 3b and 3f (local strain slightly affects the moiré periods in the three directions, but has small effect on the atomic structures of different stacking regions). Our measurement further indicates that the different stacking configurations of the MoM not only affect the moiré patterns in the STM images, but also strongly influence the spatial distributions of the flat bands in the TTG. The left panels of Figs. 3c and 3g show typical dI/dV spectra recorded in the AAA and AAA-BR regions of the samples 2 and 3, respectively. The low-energy pronounced peaks in the spectra are attributed to the high density of states (DOS) of the flat bands in the TTG. Our theoretical calculations, considering the twist angle pairs determined in our experiment, confirm the existence of flat bands, which result in high local DOS near the Fermi level (Figs. 3d and 3h). The right panels of Figs. 3c and 3g display the measured energy-fixed STS, i.e., dI/dV, maps over several moiré unit cells, which directly reflect the spatial distributions of the flat bands. In the central area of the AA-MoM region, the flat bands are concentrated on the AAA stacking region and the $C_6$ symmetry of electronic states is preserved, whereas, in the area away from the center of the AA-MoM region, the flat bands are distributed in the AAA, DW-AA-1 and DW-AA-3 stacking regions. Thus, in these regions, the $C_6$ symmetry of electronic states is broken. In Figs. 3d and 3h, we project the theoretical DOS of the flat bands onto the top graphene layer, where the electronic states are mainly detected in the STM measurements. The distribution in Fig. 3g is slightly different to that in Fig. 3h, but it is much similar to local DOS on the middle layer (see Fig. S10). It implies that the experimental feature has some contributions from the middle layer. Our theoretical results effectively capture the key features observed in our experiment.

Our STS measurements indicates that all the studied TTG samples in this work exhibit pronounced low-energy peaks in the spectra (see Figure 3 and Extended Fig. 1, 3-5). However, the full width at half maximum (FWHM) of these low-energy DOS peaks depends sensitively on the twist angle pair (see Extended Fig. 7). Figure 4a shows representative STS spectra recorded in the sample 1 with the twisted angle pair (2.85°, -1.9°). We notice that, the FWHM of the low energy DOS peaks (97 meV and 86 meV) are larger than that in the sample 2 (63 meV) and the sample 3 (51 meV). Note that the fitting results of the FWHM in the experiment can be influenced by the temperature, gating effect induced by tip[36], and so on (we only compare the FWHM of the peaks having the signal mainly from the topmost graphene layer). To further understand our experimental results, we calculated the bandwidth of the low-energy electronic states of the TTG systems with different angle pairs. Figures 4b and 4c show continuum results of the TTG with ($\theta_{TM}$, $\theta_{MB}$) for {m = 1, n = -2} and {m = 3, n = -2}, where is clear that the bandwidths strongly depend on the values of m, n and $\theta_0$. Our experiment, supported by theoretical calculations, indicates that TTG can exhibit robust flat bands at an infinite set of magic-angle pairs. Moreover, there are several degrees of freedom to tune the bandwidth, a crucial aspect for realizing novel correlated phases.

Another significant finding related to the flat bands is their distinct layer distributions in the TTG, depending on the twist angle pair. In Figure 4a, the two peaks flanking the Fermi level in the representative spectra are noticeably weaker compared to the calculated results. Meanwhile, Figure 4d shows the STS map of the DOS peak at 40 meV in sample 1, which follows the period of $\theta_0$ but with a very weak signal. To understand the reason for this observed weak signal, we simulated the distribution of this LDOS peak on each layer. As shown in Fig. 4d, the LDOS is mainly concentrated on the middle and bottom layers, while the other two strong LDOS peaks are primarily concentrated on the top and middle layers (see Fig. S9). This result implies that the electronic properties of sample 1 can be understood as a superposition of two TBG. To gain a comprehensive understanding of this, we calculated the DOS of the sample 1 and the DOS of two TBG with twist angles 2.85° and 1.9° respectively, as shown in Fig. 4f. For comparison, we also calculated the DOS of the sample 3 and the DOS of two TBG with twisted angles 0.88° and 1.76° respectively, as shown in Fig. 4e. It is noticeable that the DOS of the sample 1 is nearly identical to the superposition of that of the two TBG. This implies that the electronic properties of the sample 1 can be understood as a superposition of the two TBG and the coupling between θ$_{TM}$ and θ$_{MB}$ is rather weak. In contrast, our calculation, supported by our experiment, indicates that the flat bands of sample 2 and sample 3 have nearly equal contributions from the three twisted graphene layers, and the coupling between θ$_{TM}$ and θ$_{MB}$ is relatively strong (see Fig. S1).

In our experiment, the observed structure reconstruction is significantly different from the case of $\theta_{TM}/\theta_{MB} \approx \pm 1$. In the studied five TTG samples, the supermoiré scale ($\sim l_{MoM}$) lattice relaxation is weak and does not result in large-area single-moiré structure, as observed in the case of $\theta_{TM}/\theta_{MB} \approx \pm 1$[29,30,31,37]. Instead, the observed lattice relaxations behave similar to the TBG case[38-41]. For instance, when one of the twist angles in the angle pair is small (< ~ 1°), the AAA regions shrink and the ABA regions expand to a triangular domain separated by sharp domain walls (Fig. 3). Similarly, our theoretical calculation indicates that the lattice relaxation renormalizes the Fermi velocity of the narrow bands (Figs. S1-S3). The slight difference is that the lattice relaxation effect in the TTG is not only dependent on the magnitude of the twist angles, but also on the ratio $\theta_{TM}/\theta_{MB}$. Up to now, the structure reconstruction of the TTG is investigated only in few angle pairs and how it behaves in other angle pairs is still unclear.

To gain insights into the electronic structure, we complement our TB calculations with a continuum model to perform a detailed analysis of the electronic properties of all samples (see Supplementary Note 4). We notice that there is a slight difference between the DOS obtained with the TB calculations and the dI/dV spectra. We believe that this discrepancy is related to the absence of band renormalization due to electron-electron interactions in the TB calculations. In fact, our theoretical analysis reveals that all combinations exhibit narrow bands in the middle of the spectrum whose bandwidth is comparable with the effective long-range Coulomb interaction strength. Therefore, we can estimate the strength of electron-electron interactions and obtain the filling-dependent renormalized bands through a self-consistent Hartree potential (see Supplementary Note 4). Focusing solely on samples featuring well-defined and isolated narrow bands, we observed electronic band renormalization with filling and a pinning of the Fermi energy to the van Hove singularity, closely resembling the effects observed in TBG. Our findings indicate that, contrary to helical TTG where the interaction-induced band renormalization is weak[42], in our samples the electrostatic effects appear to be significant. On the other hand, we also found that in order to properly describe the electrostatic interactions in our systems, we require a Hartree potential whose Fourier expansion is beyond a first harmonic approximation (see Supplementary Note 5). The additional Fourier components indicate a quasicrystal structure, which is also reflected in the real-space shape of the Hartree potential (see Supplementary Note 5). Therefore, the presence of a strong Hartree potential and its dependence on doping suggests that electrostatic interactions play a significant role in these systems.

In summary, the structural and electronic characteristics of the TTG are carefully studied via STM and STS measurements, complemented with a direct comparison with theory. We demonstrate the existence of robust flat bands in TTG with different twist angle pairs, identifying the theoretically predicted magic phases. Specially, the flat bands in this system are determined by the quasiperiodicity, in the scale of several nanometres, arising from incommensurability between two moiré lattices. Our experiment indicates that TTG is a powerful platform to explore the strongly correlated physics embedded in the flat bands of quasicrystals.

**References**

1. Balents, L., Dean, C. R., Efetov, D. K. & Young, A. F. Superconductivity and strong correlations in moiré flat bands. *Nat. Phys.* **16**, 725-733 (2020).

2. Andrei, E. Y. & MacDonald, A. H. Graphene bilayers with a twist. *Nat. Mater.* **19**, 1265-1275 (2020).

3. Ren, Y.-N., Zhang, Y., Liu, Y.-W. & He, L. Twistronics in graphene-based van der Waals structures. *Chin. Phys. B* **29**, 117303 (2020).

4. Bistritzer, R. & MacDonald, A. H. Moiré bands in twisted double-layer graphene. *Proceedings of the National Academy of Sciences* **108**, 12233–12237 (2011).

5. Yin, L.-J., Qiao, J.-B., Zuo, W. J., Li, W. T. & He, L. Experimental evidence for non-Abelian gauge potentials in twisted graphene bilayers. *Phys. Rev. B* **92**, 081406(R) (2015).

6. Cao, Y. *et al.* Correlated insulator behaviour at half-filling in magic-angle graphene superlattices. *Nature* **556**, 80–84 (2018).

7. Cao, Y. *et al.* Unconventional superconductivity in magic-angle graphene superlattices. *Nature* **556**, 43–50 (2018).

8. Lopes dos Santos, J. M. B., Peres, N. M. R. & Castro Neto, A. H. Graphene Bilayer with a Twist: Electronic Structure. *Phys. Rev. Lett.* **99**, 256802 (2007).

9. Serlin, M. *et al.* Intrinsic quantized anomalous Hall effect in a moiré heterostructure. *Science* **367**, 900–903 (2020).

10. Hao, Z. *et al.* Electric field–tunable superconductivity in alternating-twist magic-angle trilayer graphene. *Science* **371**, 1133–1138 (2021).

11. Park, J. M., Cao, Y., Watanabe, K., Taniguchi, T. & Jarillo-Herrero, P. Tunable strongly coupled superconductivity in magic-angle twisted trilayer graphene. *Nature* **590**, 249–255 (2021).

12. Xu, S. *et al.* Tunable van Hove singularities and correlated states in twisted monolayer–bilayer graphene. *Nat. Phys.* **17**, 619–626 (2021).

13. Chen, S. *et al.* Electrically tunable correlated and topological states in twisted monolayer–bilayer graphene. *Nat. Phys.* **17**, 374–380 (2021).

14. Polshyn, H. *et al.* Electrical switching of magnetic order in an orbital Chern insulator. *Nature* **588**, 66–70 (2020).

15. Cao, Y. *et al.* Tunable correlated states and spin-polarized phases in twisted bilayer–bilayer graphene. *Nature* **583**, 215–220 (2020).

16. Liu, X. *et al.* Tunable spin-polarized correlated states in twisted double bilayer graphene. *Nature* **583**, 221–225 (2020).

17. Zheng, Z. *et al.* Unconventional ferroelectricity in moiré heterostructures. *Nature* **588**, 71–76 (2020).

18. Chen, G. *et al.* Signatures of tunable superconductivity in a trilayer graphene moiré superlattice. *Nature* **572**, 215–219 (2019).

19. Li, H. *et al.* Imaging two-dimensional generalized Wigner crystals. *Nature* **597**, 650–654 (2021).

20. Regan, E. C. *et al.* Mott and generalized Wigner crystal states in $WSe_2/WS_2$ moiré superlattices. *Nature* **579**, 359–363 (2020).

21. Tang, Y. *et al.* Simulation of Hubbard model physics in $WSe_2/WS_2$ moiré superlattices. *Nature* **579**, 353–358 (2020).

22. Jin, C. *et al.* Stripe phases in $WSe_2/WS_2$ moiré superlattices. *Nat. Mater.* **20**, 940–944 (2021).

23. Zhang, Z. *et al.* Flat bands in twisted bilayer transition metal dichalcogenides. *Nat. Phys.* **16**, 1093–1096 (2020).

24. Cao, Y., Park, J. M., Watanabe, K., Taniguchi, T. & Jarillo-Herrero, P. Pauli-limit violation and re-entrant superconductivity in moiré graphene. *Nature* **595**, 526–531 (2021).

25. Uri, A. *et al.* Superconductivity and strong interactions in a tunable moiré quasicrystal. *Nature* **620**, 762–767 (2023).

26. Kim, H. *et al.* Evidence for unconventional superconductivity in twisted trilayer graphene. *Nature* **606**, 494–500 (2022).

27. Liu, X., Zhang, N. J., Watanabe, K., Taniguchi, T. & Li, J. I. A. Isospin order in superconducting magic-angle twisted trilayer graphene. *Nat. Phys.* **18**, 522–527 (2022).

28. Shen, C. *et al.* Dirac spectroscopy of strongly correlated phases in twisted trilayer graphene. *Nat. Mater.* **22**, 316–321 (2023).

29. Yang, C., May-Mann, J., Zhu, Z. & Devakul, T. Multi-moir\'{e} trilayer graphene: lattice relaxation, electronic structure, and magic angles. Preprint at https://doi.org/10.48550/arXiv.2310.12961 (2023).

30. Nakatsuji, N., Kawakami, T. & Koshino, M. Multiscale Lattice Relaxation in General Twisted Trilayer Graphenes. *Phys. Rev. X* **13**, 041007 (2023).

31. Turkel, S. *et al.* Orderly disorder in magic-angle twisted trilayer graphene. *Science* **376**, 193–199 (2022).

32. Foo, D. C. W. *et al.* Extended magic phase in twisted graphene multilayers. Preprint at https://doi.org/10.48550/arXiv.2305.18080 (2023).

33. Popov, F. K. & Tarnopolsky, G. Magic Angle Butterfly in Twisted Trilayer Graphene. Preprint at https://doi.org/10.48550/arXiv.2305.16385 (2023).

34. Devakul, T. *et al.* Magic-angle helical trilayer graphene. *Science Advances* **9**, eadi6063 (2023).

35. Xia, L.-Q. *et al.* Helical trilayer graphene: a moir\'e platform for strongly-interacting topological bands. Preprint at https://doi.org/10.48550/arXiv.2310.12204 (2023).

36. Xie, Y. *et al.* Spectroscopic signatures of many-body correlations in magic-angle twisted bilayer graphene. *Nature* **572**, 101–105 (2019).

37. Craig, I. M. *et al.* Local atomic stacking and symmetry in twisted graphene trilayers. *Nat. Mater.* 1–8 (2024) doi:10.1038/s41563-023-01783-y.

38. Zheng, Q. *et al.* Tunable sample-wide electronic Kagome lattice in low-angle twisted bilayer graphene. *Phys. Rev. Lett.* **129**, 076803 (2022).

39. Ren, Y.-N. *et al.* Real-space mapping of local sub-degree lattice rotations in twisted bilayer graphene magnified by moiré superlattices. *Nano Lett.* **23**, 1836 (2023).

40. Hao, C.-R. *et al.* Creating Custom-designed Moiré Magnifying Glass to Probe Local Atomic Lattice Rotations in Twisted Bilayer Graphene. *Phys. Rev. B.* **108**, 125429 (2023).

41. Zhou, X.-F. *et al.* Coexistence of reconstructed and unreconstructed structures in structural transition regime of twisted bilayer graphene. *Phys. Rev. B.* **107**, 125410 (2023).

42. Kwan, Y. H., Ledwith, P. J., Lo, C. F. B. & Devakul, T. Strong-coupling topological states and phase transitions in helical trilayer graphene. Preprint at https://doi.org/10.48550/arXiv.2308.09706 (2023).

43. Kim, K. *et al.* van der Waals Heterostructures with High Accuracy Rotational Alignment. *Nano Lett.* **16**, 1989–1995 (2016).

44. Yu, G., Wu, Z., Zhan, Z., Katsnelson, M. I. & Yuan, S. Dodecagonal bilayer graphene quasicrystal and its approximants. npj Comput Mater 5, 1–10 (2019).

45. Kolmogorov, A. N. & Crespi, V. H. Registry-dependent interlayer potential for graphitic systems. *Phys. Rev. B* **71**, 235415 (2005).

46. Los, J. H., Ghiringhelli, L. M., Meijer, E. J. & Fasolino, A. Improved long-range reactive bond-order potential for carbon. I. Construction. *Phys. Rev. B* **72**, 214102 (2005).

47. Plimpton, S. Fast Parallel Algorithms for Short-Range Molecular Dynamics. *Journal of Computational Physics* **117**, 1–19 (1995).

48. Li, Y., Zhan, Z., Kuang, X., Li, Y. & Yuan, S. TBPLaS: A tight-binding package for large-scale simulation. Computer Physics Communications 285, 108632 (2023).

49. Cea, T., Walet, N. R. & Guinea, F. Electronic band structure and pinning of Fermi energy to Van Hove singularities in twisted bilayer graphene: A self-consistent approach. Phys. Rev. B 100, 205113 (2019).

**Table 1** | The detailed information about the five samples in our experiment.

|  | m | $\theta_{TM}$(°) | $\lambda_{TM}$(nm) | n | $\theta_{MB}$(°) | $\lambda_{MB}$(nm) | $\theta_0$(°) | $l_{\theta_0}$(nm) | $l_{MoM}$(nm) |
|---|---|---|---|---|---|---|---|---|---|
| **Sample 1** | 3 | 2.85 | 4.95 | -2 | -1.9 | 7.42 | 0.95 | 14.84 | 894.9 |
| **Sample 2** | 1 | 0.54 | 26.11 | -2 | -1.08 | 13.05 | 0.54 | 26.11 | 2769.5 |
| **Sample 3** | 1 | 0.88 | 16.02 | -2 | -1.76 | 8.01 | 0.88 | 16.02 | 1042.9 |
| **Sample 4** | 2 | 0.92 | 15.32 | -3 | -1.38 | 10.21 | 0.46 | 30.64 | 3816.6 |
| **Sample 5** | 7 | 2.91 | 4.85 | -3 | -1.25 | 11.32 | 0.415 | 33.98 | 4689.1 |

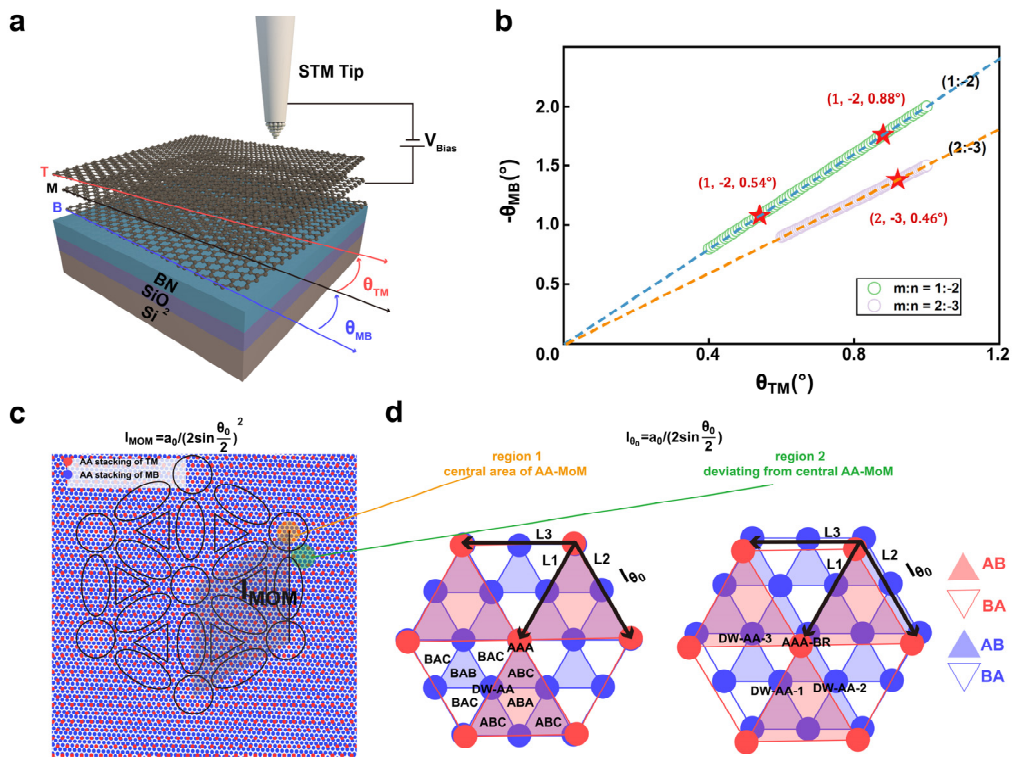

**Fig. 1 | Schematic illustration of the MoM and calculated magic angle pairs ($\theta_{TM}$, $\theta_{MB}$) in TTG. a,** Schematic of the STM set-up of the TTG device. **b,** Calculated magic angle pairs ($\theta_{TM}$, $\theta_{MB}$) for TTG with {1, -2} and {2, -3}. The magic angle pairs measured in our experiment are marked by red stars. **c,** Schematic of the MoM pattern in TTG with {1, -2}. The black quadrangle represents the unit cell of the MoM pattern. The circles represent the AA-MoM, the triangles represent the AB/BA-MoM, and the ovals represent the DW-MoM. **d,** Local structures of the $l_{\theta_0}$ pattern in the central area of AA-MoM (region 1) and area deviating from AA-MoM (region 2) of the TTG. The filled circles, filled triangles, and empty triangles indicate AA, AB, and BA stacking of individual moiré patterns respectively. Obviously, the stacking configurations of DW-AA along the three directions are the same in region 1, but they are different in region

2, which are labeled as DW-AA-1, DW-AA-2 and DW-AA-3, respectively. Therefore, the $C_6$ rotation symmetry is preserved in region 1 but broken in region 2.

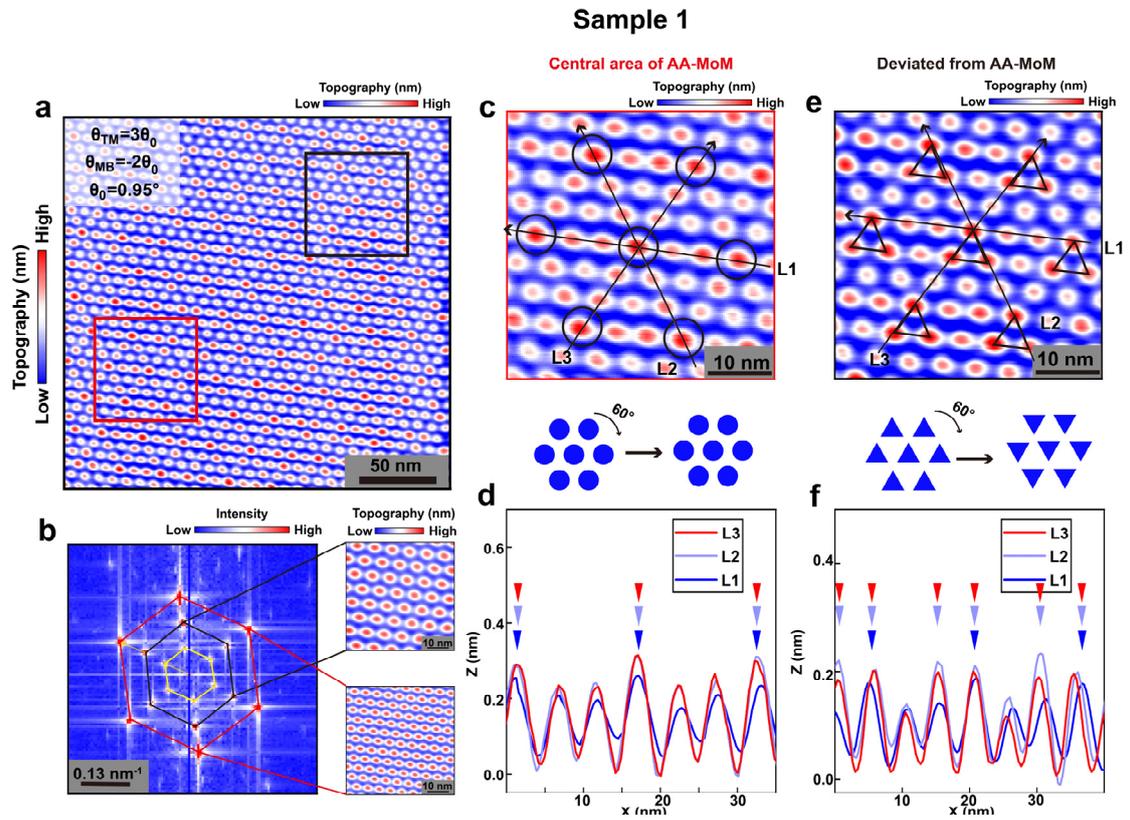

**Fig. 2 | Local characterization of stacking configurations in a TTG. a,** STM topography of sample 1 ($V_{bias}$ = 0.6 V, I = 100 pA). **b,** Left panel: FFT image of panel a. The hexagons marked the reciprocal moiré lattices of $\theta_{TM}$ (red), $\theta_{MB}$ (black) and $\theta_0$ (yellow). Right panel: Fourier-filtered images based on the first-order moiré spots of $\theta_{TM}$ and $\theta_{MB}$ in the left panel. **c, e,** Upper panels: Zoom-in images of red and black frames in panel a, representing the central area of AA-MoM and the region deviating from AA-MoM, respectively. The circles represent the AAA stacking while triangles represent the AAA-BR stacking. Bottom panels: the simplified representation of the circles and triangles in the upper panel and the arrangement after rotating 60°. The $C_6$

rotation symmetry in panel c is preserved while it is broken in panel e. **d, f,** Height profile of the lines along the three different directions in panel c and e, respectively. Inverted triangles in different colors refer to the AAA or AAA-BR stackings along the three different directions. The different arrangements of stackings along the three directions also reveal the breaking of the $C_6$ rotation symmetry.

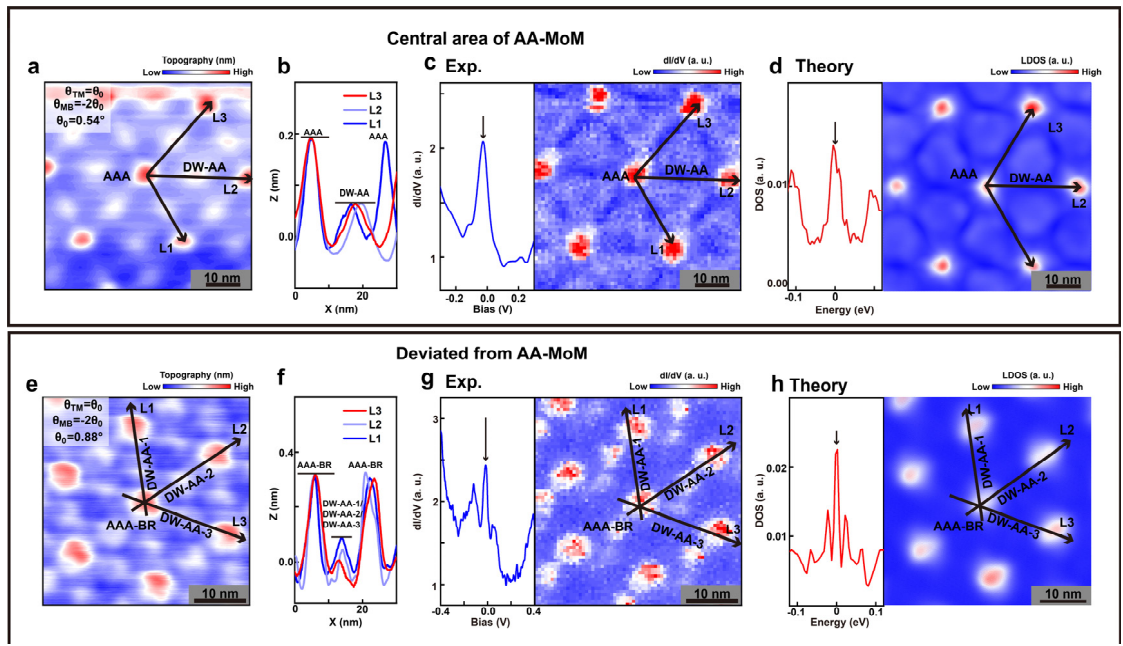

**Fig. 3 | Flat bands and their spatial distributions in the TTGs. a, e,** STM topographies of sample 2 ($V_{bias}$ = 0.4 V, I = 300 pA) and sample 3 ($V_{bias}$ = −0.8 V, I = 300 pA). **b, f,** Height profile lines along the three different directions in panels a and e, respectively. From panel b, the height differences between AAA and DW-AA stackings along the three directions are almost the same. In panel f, the height differences between AAA-BR and DW-AA along three directions are quite different, because of the different stacking configurations, labeled as DW-AA-1, DW-AA-2 and DW-AA-3. **c, g,** Left panel: dI/dV spectra recorded in AAA and AAA-BR stackings, respectively. Right

panel: dI/dV maps of area in panel a and e with the fixed sample bias, -20 and -50 meV, respectively. **d, h,** Right panels: Tight-binding calculations of the DOS of two TTGs with ($\theta_{TM}=\theta_0, \theta_{MB}=-2\theta_0$ and $\theta_0=0.54°$) and ($\theta_{TM}=\theta_0, \theta_{MB}=-2\theta_0$ and $\theta_0=0.88°$) respectively. Right panels: Calculated DOS on the top layer of the two TTGs at E = 0 eV.

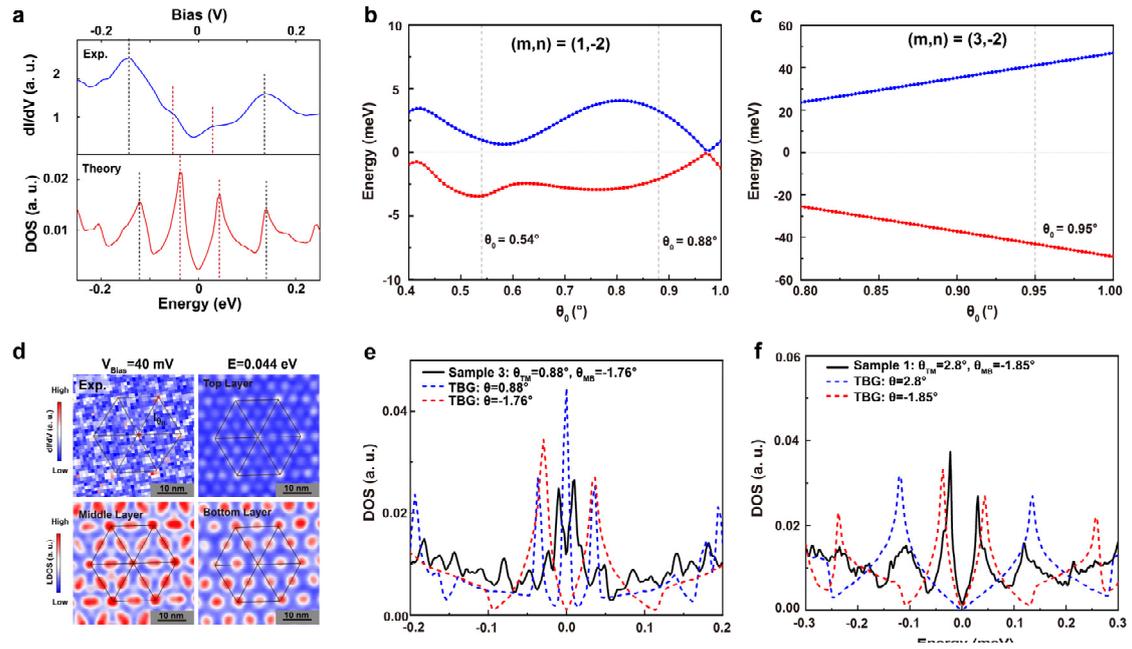

**Fig. 4 | Electronic properties of TTGs with various twist angle pairs. a,** Upper panel: dI/dV spectra recorded in AAA stacking region of the sample 1. Lower panel: Tight-binding calculations of the DOS of a TTG with ($\theta_{TM}=3\theta_0, \theta_{MB}=-2\theta_0$ and $\theta_0=0.95°$). **b, c,** Energy average of K points of the band edges (valence band (red), conduction band (blue)) in the BZ for different $\theta_0$ with angle pairs (m,n) = (1,-2) and (3,-2). The dotted lines represent the positions of three samples measured in this work. The samples 2 ($\theta_0 = 0.54°$) and 3 ($\theta_0 = 0.88°$) have narrow bands near the charge neutrality point (CNP), while the width of the band near the CNP is beyond 40 meV for sample 1 ($\theta_0 = 0.95°$). **d,** Upper left panel: dI/dV maps with the fixed sample bias, 40 mV. Other panels:

Calculated DOS on the top, middle, bottom layers of the sample 1 at E = 0.044 eV. **e,** The DOS of sample 3 (black solid line) and TBG with θ = 0.88° (blue dashed line) and θ = -1.76° (red dashed line). **f,** The DOS of sample 1 (black solid line), and of the TBG with θ = 2.8° (blue dashed line) and θ = -1.85° (red dashed line). For simplicity, we compare the DOS of the moiré structures without lattice relaxations. The number and location of peaks in sample 1 are the same to the combination of the two TBG, which indicates the weak coupling strength of the two moiré patterns. The number and location of peaks in sample 3 are totally different to the combination of the two TBG, which means the strong coupling strength of the two moiré patterns.

## Methods

### Device fabrication

The device was fabricated by using the "tear and stack" method[43]. A thick h-BN was picked up by Polydimethylsiloxane (PDMS) film while the graphene exfoliated on SiO2/Si chips was pre-cut into three parts by atomic force microscopy (AFM) tip. Next, the three parts were picked by the thick h-BN piece by piece with manually adjusted twisted angles. Finally, with the help of another PDMS, the whole heterostructure was upturned and transferred to the SiO$_2$/Si chip with pre-coated Au/Cr electrode and the graphite flake was used to connect the electrode with the sample.

### STM measurements

STM/STS measurements were performed in low-temperature (78 K) and ultrahigh-vacuum (~$10^{-10}$ Torr) scanning probe microscopes [USM-1400] from UNISOKU. The tips were obtained by chemical etching from a tungsten wire. The differential conductance (d$I$/d$V$) measurements were taken by a standard lock-in technique with an ac bias modulation of 5 mV and 793 Hz signal added to the tunneling bias and performed in a constant-current mode, where the current feedback was left on while the bias voltage changed allowing the tip to change its height. Electrochemically etched tungsten tips were used for imaging and spectroscopy.

**The density of states and charge distribution calculations**

We employed an atomic tight-binding (TB) model to compute the density of states and LDOS map of the TTG (detailed in Supplementary Note 1-2). We used a round disk method to construct the TTG with commensurate or incommensurate twist angles[44]. Then, the samples were relaxed with the semi-classical molecular dynamics implemented in LAMMPS[45-47]. The radius of the disk was $700a$, which contained 10 million carbon atoms. To obtain the DOS and LDOS map of these complex samples, we utilized the tight-binding propagation method without diagonalization process, which is implemented in the TBPLaS package[48].

**Continuum model and Hartree calculations**

We determine the electronic band structure of all samples employing a low-energy continuum model. The methodology outlined in ref.[32] is followed, considering an approximately commensurate trilayer system (detailed in Supplementary Note 4). A self-consistent Hartree calculation of the electronic bands is carried out[49] specifically for samples 2 and 3, because they are characterized by well-isolated narrow bands (detailed in Supplementary Note 4).

## Data availability

Source data are provided with this paper. All other data that support the plots within this paper and other findings of this study are available from the corresponding authors on reasonable request.

## Acknowledgments

This work was supported by the National Key R and D Program of China (Grant Nos. 2021YFA1401900 and 2021YFA1400100), National Natural Science Foundation of China (Grant Nos. 12141401, 11974050), and "the Fundamental Research Funds for the Central Universities". Z.Z. acknowledges support funding from the European Union's Horizon 2020 research and innovation programme under the Marie Skłodowska-Curie grant agreement No 101034431 P.A.P, F.G and ZZ. acknowledge support from NOVMOMAT, Grant PID2022-142162NB-I00 funded by MCIN/AEI/ 10.13039/501100011033 , by "ERDF A way of making Europe" and from the "Severo Ochoa" Programme for Centres of Excellence in R&D (CEX2020-001039-S / AEI / 10.13039/501100011033). We thank the Core Facility of Wuhan University for providing the computational resources. The devices were fabricated using the transfer platform from Shanghai Onway Technology Co. Ltd.

## Author contributions

C.-Y.H. fabricated the device. C.-Y.H., J.-Q.H. and Y.-X.Z. performed the measurements. C.-Y.H., Z.Z., F.G., and L.H. analyzed the data. Z.Z., and P.A.P. provided theoretical analysis, supervised by F.G. K.W. and T.T. provided the hexagonal boron nitride crystals. C.-Y.H., Z.Z., P.A.P. and L.H. wrote the manuscript with input from all co-authors.

## Competing interests

The authors declare no competing financial interests.